# On the statistical correlations of random polarization and electric depolarization fields in ferroelectrics

Yuri A. Genenko[1], Olga Y. Mazur[2,3], and Leonid I. Stefanovich[3]

[1]Institute of Materials Science, Technical University of Darmstadt, Otto-Berndt-Str, 3, 64287 Darmstadt, Germany

[2]Technical University of Liberec, Studentska 1402/2, 46117 Liberec, Czech Republic

[3]Branch for Physics of Mining Processes of the M.S. Poliakov Institute of Geotechnical Mechanics of the National Academy of Sciences of Ukraine, 49600 Dnipro, Simferopolska st., 2a, Ukraine

Abstract

A conceptual problem of the electric-field mediated polarization correlations during a stochastic formation of polarization domain structure in ferroelectrics is addressed by using an exactly solvable stochastic model of polarization development in a uniaxial ferroelectric [Phys. Rev. B 107, 144109 (2023)]. A full set of time-dependent two-point correlation coefficients between all random variables is derived analytically, evaluated numerically and presented graphically in 3D. They are particularly required for the analysis of nonlinear phenomena involving spatial dispersion like optical second harmonic generation and scattering.

## 1. Introduction

Electric depolarization fields play an important role in the formation of domain structures in ferroelectrics. They contribute essentially into the system thermodynamics leading to the formation of polarization domain structures with a minimum energy which often have rather regular patterns well correlated over large distances [1,2]. There is, however, a conceptual problem with the understanding of the depolarization field role in the polarization switching when a system is subject to a high enough electric field. The Ishibashi stochastic theory of the polarization reversal, induced by an opposite electric field applied to a uniformly polarized system, assumes statistically independent nucleation and growth of domains of opposite direction and thus neglects a possible interaction of these domains [3]. Indeed, this theory, usually called the Kolmogorov-Avrami-Ishibashi (KAI) model, accounting for one-component polarization reversals and thus applicable only for 180°-switching processes, adopts a concept developed by Kolmogorov for the problem of solidification from a melt [4], where the reversed domains induce no physical fields. The situation in ferroelectrics, which are often also ferroelastics, is different since the reversed domains induce long-range electric and elastic fields and are expected to interact with each other. Despite this drawback the concept of statistically independent nucleation and growth of reversed polarization domains, ignoring their electric or elastic interactions, was quite successfully applied to different single-crystals and further extended in numerous works [5-9]. The application of this approach to polycrystalline ferroelectrics turned out to be less successful which was, however, improved by introducing the nucleation limited switching (NLS) model assuming a hypothetical statistical distribution of local polarization switching times [10-16]. Subsequently, the distribution of switching times was derived from the statistical distribution of local electric fields in the inhomogeneous field mechanism (IFM) model [17,18] which was then successfully applied to various polycrystalline materials [19-32]. Furthermore, based on the KAI-approach the polarization reversal proceeding by sequential switching events was successfully described in tetragonal [33], rhombohedral [34] and orthorhombic [35] ferroelectrics by means of the multistep switching mechanism (MSM) model. Note that all the above mentioned (KAI, NLS, IFM and MSM) models neglected interactions of the nucleating reversed domains and their respective correlations.

In contrast to the hypotheses adopted in the above stochastic models, simulations of polarization switching in single- and polycrystalline materials, accounting electric and elastic interactions, revealed remarkable polarization correlations at different spatial scales. Thus, comprehensive phase-field simulations by Zhou at al. [36] demonstrated a strongly correlated self-organization behavior and a coherent temporal evolution of self-accommodating domains in a single crystal of $BaTiO_3$ substantially reducing depolarization fields. A similar behavior was observed in the lattice model of thermally activated dipoles [37]. Molecular dynamics simulations of domain wall motion in the $PbTiO_3$ single crystal also exhibited spatially correlated coherent switching scenarios [38]. On the other hand, simulations of the polarization reversal in polycrystalline ferroelectric materials of different crystalline symmetries by means of the self-consistent mesoscopic switching (SMS) model [39,40] disclosed only short-range correlations at the grain size scale that in principle agrees with comprehensive simulations by Indergand et al. [41].

Experimental studies confirm the presence of correlations at different spatial scales. Thus, direct observations of the polarization reversal in a single crystal of $BaTiO_3$ by polarized light microscope demonstrated a highly coherent switching process with domain walls moving in such a way as to suppress electric depolarization fields and release mechanical stresses [42,43] that confirms the phase-field simulations [36]. Similarly, in polycrystalline lead zirconate titanate (PZT) thin films, piezoresponse force microscopy (PFM) and transmission electron microscopy revealed correlations in polarization response ranging from a few grains [44] to agglomerations of $10^2$–$10^3$ grains [45,46]. The latter correlations, however, seem to be related to the elastic interaction via a substrate [47,48]. Correlations of polarization dynamics in bulk ferroelectric ceramics, studied by the grain-resolved three-dimensional x-ray diffraction, appeared to involve approximately 10–20 grains thus exhibiting a characteristic correlation scale about the grain size [49-51] that confirms the predictions of the SMS model simulations [39,40].

Finally, simulations [36-41] are in principle agreement with experimental observations demonstrating polarization correlations at different scales in single- and polycrystalline materials [42-51] but contradict the basic assumptions of stochastic models [3-35] assuming statistically independent nucleation and growth of reversed polarization domains. The effect of emerging stochastic depolarization fields on the polarization development at different locations should result, indeed, from the fundamental relation between polarization and electric field expressed by the Gauss equation. In fact, the problem of the stochastic theory is the need to take into account both temporal and spatial correlations in the process of domain formation which still presents a difficult task. In this study, we first try to comprehend the time-dependent spatial correlations of polarization and electric field during the stochastic domain development from an initial disordered state. To this end we use the exactly solvable model of stochastic domain formation in uniaxial ferroelectrics which accounts self-consistently for electric interaction between local polarizations [52]. We derive a complete set of two-point correlation functions between polarization and electric field components and between the field components themselves and show that some of these correlations vanish for general symmetry reasons in certain directions and planes but generally they are rather relevant for the domain structure development.

## 2. The model

Let us consider a uniaxial single-crystalline ferroelectric/nonferroelastic with the polarization along the $z$-axis of the Cartesian coordinate system as is shown in Fig. 1. In a typical experimental geometry of a ferroelectric plate of thickness $h_f$, attached to a bottom electrode and a dielectric layer of thickness $h_d$ at the top side covered with a top electrode allowing for application of an external field. The Landau-Ginzburg-Devonshire (LGD) energy functional of the system can be presented in the form [53,54]

$$\Phi = \Phi_0 + \int_{V_f} \left[\frac{1}{2}AP_z^2 + \frac{1}{4}BP_z^4 + \frac{1}{2}G(\nabla P_z)^2 - P_z E_z - \frac{\varepsilon_0 \varepsilon_b}{2}\boldsymbol{E}^2\right] dV - \int_{V_d} \frac{\varepsilon_0 \varepsilon_d}{2}\boldsymbol{E}^2 dV \qquad (1)$$

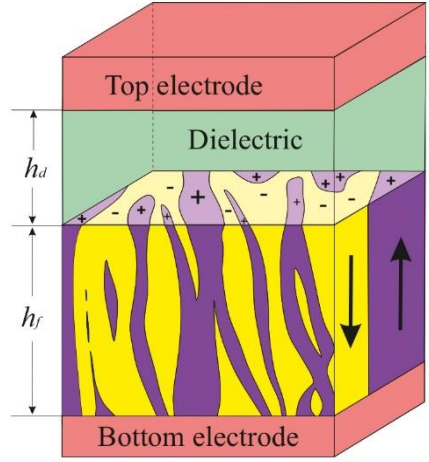

Fig. 1. Problem layout: A ferroelectric slab of thickness $h_f$, placed on a bottom electrode and separated from a top electrode by a dielectric layer of thickness $h_d$, is infinite in $(x, y)$-plane parallel to the ferroelectric surface. Polarization direction is along the vertical $z$-axis of the Cartesian $(x, y, z)$-frame.

with the coefficient $A = \alpha_0(T - T_c)$, $\alpha_0 > 0$, $T < T_c$, which is the temperature of the second order paraelectric-ferroelectric phase transition, and the other temperature independent coefficients $B > 0$ and $G > 0$. $\boldsymbol{E}$ denotes the local electric field, $\varepsilon_0$, $\varepsilon_d$ and $\varepsilon_b$ are the permittivity of vacuum, of the dielectric layer and the background permittivity of the ferroelectric, respectively, while $V_f$ and $V_d$ denote the volumes of the ferroelectric plate and the dielectric layer, respectively.

It is convenient for the following calculations to introduce dimensionless physical variables normalized to their natural characteristic magnitudes in the phase transition problem. Thus, we denote a dimensionless polarization $\pi = P_z/P_s$ normalized to the spontaneous equilibrium polarization $P_s = \sqrt{|A|/B}$, and a dimensionless electric field $\boldsymbol{\epsilon} = \boldsymbol{E}/E_0$ with the value of $E_0 = P_s|A|$. All spatial coordinates are normalized to a characteristic length $\lambda = \sqrt{G/|A|}$ being the characteristic domain wall thickness.

We consider the evolution of the system from an initial state obtained by quenching from the high temperature paraelectric phase to the ferroelectric one at temperature $T < T_c$. Since the initial state is a random one, all physical variables become random too and will be considered in this model as Gauss random fields as suggested previously [55,56]. Another source of randomness is the stochastic thermal noise which may have a crucial impact on polarization switching kinetics as was shown by Indergand et al. [41] by simulation of the polarization reversal from a uniformly polarized initial state in single- and polycrystalline samples. When considering the quenched initial conditions, it is known from experiments on single crystals that the initial disorder may be substantial depending on the initial temperature in the parent paraelectric phase, the quenching temperature and the cooling rate [57-60]. As was shown in the preceding authors' work [52], the initial disorder may dominate over the thermal noise in the temporal domain development in a wide temperature range away from the phase transition temperature, if the magnitude and the spatial scale of the initial disorder are large enough; the case considered in the current study neglecting thermal fluctuations.

The polarization can then be represented as $\pi(\mathbf{r}, \tau) = \bar{\pi}(\tau) + \xi(\mathbf{r}, \tau)$ with the dimensionless mean polarization magnitude $\bar{\pi}(\tau) = \langle P_z \rangle / P_s$, depending on the dimensionless time $\tau$, and the stochastic polarization $\xi(\mathbf{r}, \tau)$ such that $\langle \xi(\mathbf{r}, \tau) \rangle = 0$. Here the sign $\langle \ldots \rangle$ denotes statistical averaging over all possible system realizations. Then a dimensionless local electric field in the chosen sample geometry reads

$$\boldsymbol{\epsilon}(\mathbf{r}, \tau) = \boldsymbol{\epsilon}_a - \alpha_z \bar{\pi}(\tau) \hat{\mathbf{z}} - \boldsymbol{\nabla} \phi(\mathbf{r}, \tau) \qquad (2)$$

where $\boldsymbol{\epsilon}_a$ is a uniform electric field in the ferroelectric induced by a voltage applied to the electrodes [52]. The second term in Eq. (2) represents the mean depolarization field in the ferroelectric due to the average polarization $\bar{\pi}$ with the depolarization coefficient $\alpha_z = \frac{h_d}{(\varepsilon_d h_f + \varepsilon_b h_d)\varepsilon_0 |A|}$, and the last term is the contribution of the stochastic electric depolarization field due to the stochastic electric potential $\phi(\mathbf{r}, \tau)$, such that $\langle \boldsymbol{\nabla} \phi \rangle = 0$.

By variation of the energy functional (1) with respect to the polarization and the electric potential, respectively, a system of evolution equations can be derived [52],

$$\begin{cases} \frac{\partial \pi}{\partial \tau} = \Delta \pi + \pi - \pi^3 + \epsilon_z & (3a) \\ \Delta \phi = \eta \frac{\partial \pi}{\partial z} & (3b) \end{cases}$$

where the first one is the Landau-Khalatnikov kinetic equation and the second one is the Poisson equation with a dimensionless parameter $\eta = 1/(\varepsilon_0 \varepsilon_b |A|)$.

### 3. General relations for correlation functions

We introduce now correlation functions characterizing the system, the governing equations for which can be derived from Eqs. (3). These functions include two-point autocorrelation functions for the polarization, $K(\mathbf{s}, \tau) = \langle \xi(\mathbf{r}_1, \tau) \xi(\mathbf{r}_2, \tau) \rangle$, and the electric potential, $g(\mathbf{s}, \tau) = \langle \phi(\mathbf{r}_1, \tau) \phi(\mathbf{r}_2, \tau) \rangle$, with $\mathbf{s} = \mathbf{r}_1 - \mathbf{r}_2$, as well as cross-correlation functions between the polarization and electric field components, $\Psi_{xz}(\mathbf{s}, \tau) = \langle \epsilon_x(\mathbf{r}_1, \tau) \xi(\mathbf{r}_2, \tau) \rangle$, $\Psi_{yz}(\mathbf{s}, \tau) =$

$\langle \epsilon_y(\mathbf{r}_1,\tau)\xi(\mathbf{r}_2,\tau)\rangle$, $\Psi_{zz}(\mathbf{s},\tau) = \langle \epsilon_z(\mathbf{r}_1,\tau)\xi(\mathbf{r}_2,\tau)\rangle$, and between the electric field components themselves, $R_{\alpha\beta}(\mathbf{s},\tau) = \langle \frac{\partial \phi(\mathbf{r}_1,\tau)}{\partial r_{1\alpha}} \frac{\partial \phi(\mathbf{r}_2,\tau)}{\partial r_{2\beta}}\rangle$.

The knowledge of correlation functions for stochastic systems is required to evaluate any macroscopic physical quantities involving products of random variables. For example, for evaluation of the total energy (1) the knowledge of the correlation functions $K, R_{xx}, R_{yy}, R_{zz}$ and $\Psi_{zz}$ is necessary.

The correlation functions are interrelated with each other that can be shown by multiplying Eqs. (3) with different variables and consequent statistical averaging [52]. The details of the derivation are presented in Appendix A. Thus, the functions $g(\mathbf{s},\tau)$, $\Psi_{xz}(\mathbf{s},\tau)$ and $\Psi_{yz}(\mathbf{s},\tau)$ are related to $\Psi_{zz}(\mathbf{s},\tau)$ as

$$\Delta g(\mathbf{s},\tau) = \eta \Psi_{zz}(\mathbf{s},\tau), \quad \frac{\partial}{\partial s_z}\Psi_{xz}(\mathbf{s},\tau) = \frac{\partial}{\partial s_x}\Psi_{zz}(\mathbf{s},\tau), \quad \frac{\partial}{\partial s_z}\Psi_{yz}(\mathbf{s},\tau) = \frac{\partial}{\partial s_y}\Psi_{zz}(\mathbf{s},\tau), \quad (4)$$

while $\Psi_{zz}(\mathbf{s},\tau)$ is, in turn, related to $K(\mathbf{s},\tau)$,

$$\Delta \Psi_{zz}(\mathbf{s},\tau) = -\eta \frac{\partial^2}{\partial s_z^2} K(\mathbf{s},\tau). \quad (5)$$

Finally, the function $R_{\alpha\beta}(\mathbf{s},\tau)$ can be expressed through $g(\mathbf{s},\tau)$,

$$R_{\alpha\beta}(\mathbf{s},\tau) = -\frac{\partial^2 g(\mathbf{s},\tau)}{\partial s_\alpha \partial s_\beta}. \quad (6)$$

Using Fourier transforms defined as

$$K(\mathbf{s},\tau) = \frac{1}{(2\pi)^3} \int d^3q \exp(i\mathbf{qs}) \widetilde{K}(\mathbf{q},\tau), \quad (7a)$$

$$\widetilde{K}(\mathbf{q},\tau) = \int d^3s \exp(-i\mathbf{qs}) K(\mathbf{s},\tau) \quad (7b)$$

the relations (4-6) can be converted into explicit algebraic expressions via $\widetilde{K}(\mathbf{q},\tau)$,

$$\widetilde{\Psi}_{zz}(\mathbf{q},\tau) = -\eta \frac{q_z^2}{q^2}\widetilde{K}(\mathbf{q},\tau), \quad \widetilde{\Psi}_{xz}(\mathbf{q},\tau) = -\eta \frac{q_x q_z}{q^2}\widetilde{K}(\mathbf{q},\tau), \quad \widetilde{\Psi}_{yz}(\mathbf{q},\tau) = -\eta \frac{q_y q_z}{q^2}\widetilde{K}(\mathbf{q},\tau). \quad (8)$$

$$\widetilde{g}(\mathbf{q},\tau) = \eta^2 \frac{q_z^2}{q^4}\widetilde{K}(\mathbf{q},\tau), \quad \widetilde{R}_{\alpha\beta}(\mathbf{q},\tau) = \eta^2 \frac{q_z^2 q_\alpha q_\beta}{q^4}\widetilde{K}(\mathbf{q},\tau). \quad (9)$$

Thus, the problem of correlations is reduced to the finding of the function $\widetilde{K}(\mathbf{q},\tau)$ alone.

The latter correlation function results from the solution of the system of integro-differential equations for $\widetilde{K}(\mathbf{q},\tau)$ and $\bar{\pi}(\tau)$ derived previously [52],

$$\begin{cases} \frac{d\bar{\pi}}{d\tau} = \bar{\pi}\big(1 - \alpha_z - 3K(0,\tau)\big) - \bar{\pi}^3 + \epsilon_a & (10a) \\ \frac{d\widetilde{K}(\mathbf{q},\tau)}{d\tau} = 2\left[1 - 3\bar{\pi}^2(\tau) - 3K(0,\tau) - \left(q^2 + \eta\frac{q_z^2}{q^2}\right)\right]\widetilde{K}(\mathbf{q},\tau) & (10b) \end{cases}$$

By solving Eq. (10b), the function $\widetilde{K}(\mathbf{q},\tau)$ can be expressed through its initial value defined by correlations in the initial state after quenching [52],

$$\widetilde{K}(\mathbf{q},\tau) = \widetilde{K}(\mathbf{q},0)\mu(\tau)\exp\left[-2\left(q^2 + \eta\frac{q_z^2}{q^2}\right)\tau\right]. \tag{11}$$

Assuming hypothetically a Gaussian shape of the initial correlations $K(\mathbf{s},0)$ we get

$$\widetilde{K}(\mathbf{q},0) = (2\pi)^{\frac{3}{2}} K_0 r_c^3 \exp\left(-\frac{r_c^2 q^2}{2}\right) \tag{12}$$

with the Gauss parameter $r_c$ and the initial fluctuation magnitude $K_0$ [52]. This choice defines also the auxiliary function $\mu(\tau)$ in Eq. (11),

$$\mu(\tau) = \frac{2}{\sqrt{\pi}} \frac{D(\tau)}{D(0)} \left(1 + \frac{4\tau}{r_c^2}\right)^{3/2} \frac{\sqrt{2\eta\tau}}{\mathrm{erf}(\sqrt{2\eta\tau})} \tag{13}$$

where $D(\tau) = K(\mathbf{0},\tau)$ is the time-dependent variance, or dispersion, of spatial polarization fluctuations.

In terms of the Fourier transform, the characteristic correlation length $L(\tau)$ can be conveniently defined as [55]

$$L^{-2}(\tau) = \int d^3q\, q^2 \widetilde{K}(\mathbf{q},\tau) / \int d^3q\, \widetilde{K}(\mathbf{q},\tau). \tag{14}$$

This definition differs from that commonly used in experiment, which typically defines this length as the distance at which the correlation function is halved compared to its value at the origin [5,61]. However, since the system possesses the only dominating characteristic scale, the definitions in terms of the spatial correlation function and its Fourier transform may only be different by a factor of the order of unity, which can be well adjusted by the choice of the fitting parameter $r_c$. This was demonstrated by the successful comparison of $L(\tau)$ with experimental data in Ref. [52]. For the adopted Gaussian shape of the initial correlation function, Eq. (12), $L(\tau)$ can be explicitly evaluated by substituting Eq. (11) into Eq. (14) resulting in [52]

$$L(\tau) = \sqrt{(r_c^2 + 4\tau)/3}. \tag{15}$$

Since all the correlation functions (8,9) are proportional to $\widetilde{K}(\mathbf{q},\tau)$ and thus to $D(\tau)$, the latter appears to strongly affect the time evolution of all correlations in the system, particularly suppressing them asymptotically if the final state of the system is a single-domain one. The function $D(\tau)$ cannot be found analytically but results from the numerical solution of the system of nonlinear differential equations as was demonstrated previously [52]. Fortunately, the problem of correlations can be decoupled from the said numerical analysis by introducing the correlation coefficients normalized to $D(\tau)$ for all pairs of random variables. This allows analytical calculation of all correlation coefficients which will be performed in the following.

### 4. Correlation coefficients

The correlation coefficient for the polarization correlations, $C(\mathbf{s},\tau) = K(\mathbf{s},\tau)/D(\tau)$, was derived in Ref. [52]. The other correlation coefficients result from the respective correlation functions normalized to the products of the standard deviations of the involved variables as follows,

$$\psi_{xz}(\mathbf{s},\tau) = \frac{\Psi_{xz}(\mathbf{s},\tau)}{\sqrt{R_{xx}(0,\tau)}\sqrt{D(\tau)}}, \quad \psi_{yz}(\mathbf{s},\tau) = \frac{\Psi_{yz}(\mathbf{s},\tau)}{\sqrt{R_{yy}(0,\tau)}\sqrt{D(\tau)}}, \quad \psi_{zz}(\mathbf{s},\tau) = \frac{\Psi_{zz}(\mathbf{s},\tau)}{\sqrt{R_{zz}(0,\tau)}\sqrt{D(\tau)}} \quad (16)$$

$$r_{xx}(\mathbf{s},\tau) = \frac{R_{xx}(\mathbf{s},\tau)}{R_{xx}(0,\tau)}, \quad r_{yy}(\mathbf{s},\tau) = \frac{R_{yy}(\mathbf{s},\tau)}{R_{yy}(0,\tau)}, \quad r_{zz}(\mathbf{s},\tau) = \frac{R_{zz}(\mathbf{s},\tau)}{R_{zz}(0,\tau)}, \quad (17)$$

$$r_{xy}(\mathbf{s},\tau) = \frac{R_{xy}(\mathbf{s},\tau)}{\sqrt{R_{xx}(0,\tau)}\sqrt{R_{yy}(0,\tau)}}, \quad r_{xz}(\mathbf{s},\tau) = \frac{R_{xz}(\mathbf{s},\tau)}{\sqrt{R_{xx}(0,\tau)}\sqrt{R_{zz}(0,\tau)}}, \quad r_{yz}(\mathbf{s},\tau) = \frac{R_{yz}(\mathbf{s},\tau)}{\sqrt{R_{yy}(0,\tau)}\sqrt{R_{zz}(0,\tau)}}, \quad (18)$$

Accordingly, we start with the evaluation of the variances $R_{\alpha\alpha}(0,\tau)$ required for normalization of the coefficients (16-18). By substituting Eqs. (9) and (11) into Fourier-transform (7a) one finds for $\alpha = \beta = x$

$$R_{xx}(\mathbf{s},\tau) = \frac{\eta^2 \mu(\tau)}{(2\pi)^3} \int_0^\infty dq\, q^2\, e^{-2\tau q^2} \widetilde{K}(\mathbf{q},0) \int_0^\pi d\theta \sin^3\theta \cos^2\theta\, e^{-2\eta\tau \cos^2\theta + iq s_z \cos\theta}$$

$$\times \int_0^{2\pi} d\phi \cos^2\phi\, e^{iq \sin\theta(s_x \cos\phi + s_y \sin\phi)}. \quad (19)$$

Here and below the $q_z$-axis of spherical coordinate system in $q$-space is chosen along the direction of vector $\mathbf{s}$. For $\mathbf{s}=0$ it is reduced to the variance

$$R_{xx}(0,\tau) = \frac{\eta^2 \mu(\tau)}{(2\pi)^3} \int_0^\infty dq\, q^2\, e^{-2\tau q^2} \widetilde{K}(\mathbf{q},0) \int_0^\pi d\theta \sin^3\theta \cos^2\theta\, e^{-2\eta\tau \cos^2\theta} \int_0^{2\pi} d\phi \cos^2\phi. \quad (20)$$

By substituting the formulas (12) and (13) into Eq. (20) one obtains

$$R_{xx}(0,\tau) = \frac{\eta D(\tau)}{8\tau}\left[1 - \frac{3}{4\eta\tau}\left(1 - \frac{2}{\sqrt{\pi}} \frac{\sqrt{2\eta\tau}\exp(-2\eta\tau)}{\operatorname{erf}(\sqrt{2\eta\tau})}\right)\right]. \quad (21)$$

This function seems to be singular at $\tau \to 0$ but, in fact, it is regular and has a limit of $R_{xx}(0,0) = \frac{1}{15}\eta^2 D(0)$ where $D(0) = K_0$.

Quite similar for $\alpha = \beta = y$,

$$R_{yy}(\mathbf{s}, \tau) = \frac{\eta^2 \mu(\tau)}{(2\pi)^3} \int_0^\infty dq\, q^2\, e^{-2\tau q^2} \widetilde{K}(\mathbf{q}, 0) \int_0^\pi d\theta\, \sin^3\theta \cos^2\theta\, e^{-2\eta\tau \cos^2\theta + iq\, s_z \cos\theta}$$

$$\times \int_0^{2\pi} d\phi\, \sin^2\phi\, e^{iq \sin\theta (s_x \cos\phi + s_y \sin\phi)}\,. \tag{22}$$

For $\mathbf{s}=0$ it is reduced to the variance

$$R_{yy}(0, \tau) = \frac{\eta^2 \mu(\tau)}{(2\pi)^3} \int_0^\infty dq\, q^2\, e^{-2\tau q^2} \widetilde{K}(\mathbf{q}, 0) \int_0^\pi d\theta\, \sin^3\theta \cos^2\theta\, e^{-2\eta\tau \cos^2\theta} \int_0^{2\pi} d\phi\, \sin^2\phi\,. \tag{23}$$

By substituting the formulas (12) and (13) into Eq. (23) one obtains $R_{yy}(0, \tau) = R_{xx}(0, \tau)$, which is justified for symmetry reasons.

For the case $\alpha = \beta = z$,

$$R_{zz}(\mathbf{s}, \tau) = \frac{\eta^2 \mu(\tau)}{(2\pi)^3} \int_0^\infty dq\, q^2\, e^{-2\tau q^2} \widetilde{K}(\mathbf{q}, 0) \int_0^\pi d\theta\, \sin\theta \cos^4\theta\, e^{-2\eta\tau \cos^2\theta + iq\, s_z \cos\theta}$$

$$\times \int_0^{2\pi} d\phi\, e^{iq \sin\theta (s_x \cos\phi + s_y \sin\phi)}\,. \tag{24}$$

For $\mathbf{s}=0$ it is reduced to the variance

$$R_{zz}(0, \tau) = \frac{\eta^2 \mu(\tau)}{(2\pi)^2} \int_0^\infty dq\, q^2\, e^{-2\tau q^2} \widetilde{K}(\mathbf{q}, 0) \int_0^\pi d\theta\, \sin\theta \cos^4\theta\, e^{-2\eta\tau \cos^2\theta}\,. \tag{25}$$

By substituting the formulas (12) and (13) into Eq. (25) and integration one obtains

$$R_{zz}(0, \tau) = \frac{3 D(\tau)}{(4\tau)^2}\left[1 - \frac{2}{\sqrt{\pi}}\frac{\sqrt{2\eta\tau}\,\exp(-2\eta\tau)}{\mathrm{erf}(\sqrt{2\eta\tau})}\left(1 + \frac{4\eta\tau}{3}\right)\right]. \tag{26}$$

This function at first glance seems to be singular when $\tau \to 0$, but, in fact, it is regular with the limit of $R_{zz}(0,0) = \frac{1}{5}\eta^2 D(0)$.

### 4.1 Cylindrically symmetrical correlation coefficients

Thanks to the cylindrical symmetry of the system, the orientational dependence of correlation coefficients $C(\mathbf{s}, \tau) = K(\mathbf{s}, \tau)/D(\tau)$, $r_{zz}(\mathbf{s}, \tau)$ and $\psi_{zz}(\mathbf{s}, \tau)$ may by represented in any cross-sectional plane containing the symmetry axis $z$, for example in the plane $\mathbf{s} = (s\cdot\sin(\vartheta), 0, s\cdot\cos(\vartheta))$ where $\vartheta$ is the polar angle with respect to the polarization direction $z$. Accordingly, the correlation coefficient $C(\mathbf{s}, \tau)$ takes the form

$$C(\mathbf{s}, \tau) = \frac{3\sqrt{6}}{\pi} \frac{\sqrt{2\eta\tau}}{\mathrm{erf}(\sqrt{2\eta\tau})} \int_0^\infty dk\, k^2 e^{-3k^2/2} \int_{-1}^1 du\, e^{-2\eta\tau u^2} \cos\left(\frac{kus\cos\vartheta}{L(\tau)}\right) J_0\left(\frac{k\sqrt{1-u^2}\, s \sin\vartheta}{L(\tau)}\right). \tag{27}$$

using the correlation length $L(\tau)$, Eq. (15).

The evolution of $C(\mathbf{s},\tau)$ in time can be captured when considering a series of polar plots depending on the polar angle $\vartheta$ at the surface $s=$ const for different time moments. We present in Fig. 2 the development of $C(\mathbf{s},\tau)$ at $s=3$, assuming here and below parameters $r_c=1, \eta=1$, for a range of times $\tau$. At the early stage (*a*), when the correlation length $L(\tau)$ is smaller than the radius of the test sphere $s=3$, the polarization correlations are weak and slightly anisotropic in favor of the polarization direction ($\vartheta=0$) because randomly chosen points at the distance $s$ belong to different small domains far away from each other. At the intermediate stage (*b-d*), when $L(\tau)$ gradually approaches the radius of the test sphere, correlations are getting noticeable and distinctly anisotropic because many randomly chosen points appear within the same domain. Eventually, when the radius $L(\tau)$ reaches the value about $s$ (*e*) or completely comprises the test sphere (*f*), correlations become strong but much less anisotropic because statistically the most of randomly chosen pairs of points are within the same domain where polarization behaves coherently.

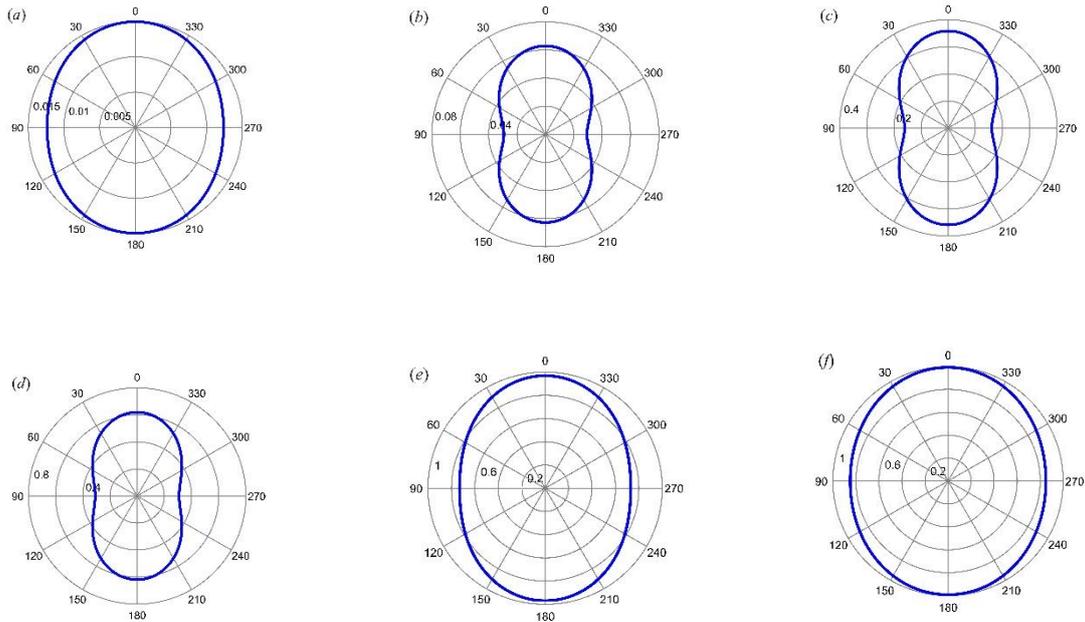

Fig. 2. Dependence of the correlation coefficient $C(\mathbf{s},\tau)$ at the distance $s=3$ in the cross section $(s_x, 0, s_z)$ on the polar angle $\vartheta$ with respect to the polar axis $z$ for the moments $\tau = 0.01\ (a), 0.1\ (b), 0.5\ (c), 1\ (d),\ 5\ (e), 10\ (f)$.

Also interesting is to observe variation of $C(\mathbf{s},\tau)$ with the distance $s$ at a fixed time. Fig. 3 demonstrates polar plots at time $\tau=1$ for different $s$. While inside the region of the size $L(\tau)$ correlations are strong and virtually isotropic (*a*), they become weaker but strongly anisotropic at distances $s \gg L(\tau)$ (*b,c*) that seems to be in favor of prevailed formation of longitudinal domains along the axis $z$.

Considering the analytical expressions for longitudinal, $C_\parallel(s_z,\tau)$, and transverse, $C_\perp(\mathbf{s}_\perp,\tau)$, correlations [52], depending essentially on the combined variable $s^2/(12L^2(\tau))$, the correlation coefficient can be well approximated as

$$C(\mathbf{s},\tau) = C_\perp(s,\tau)\sin^2\vartheta + C_\parallel(s,\tau)\cos^2\vartheta \tag{28}$$

for the distances $s^2/(12L^2(\tau)) < 1$ that is shown by dashed lines in Fig. 3

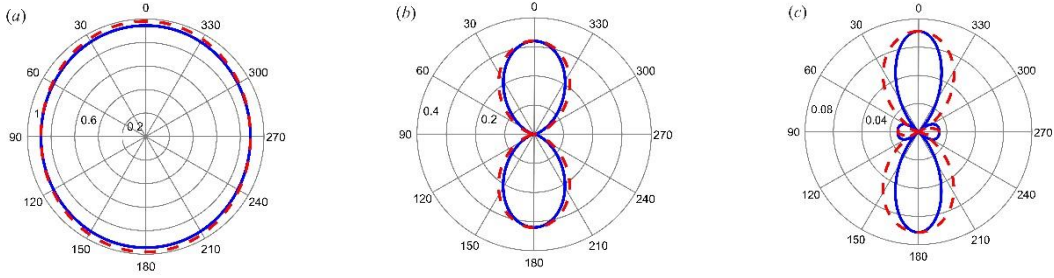

Fig. 3. Dependence of the correlation coefficient $C(\mathbf{s},\tau)$ at the time $\tau = 1$ on the polar angle $\vartheta$ with respect to the polar axis $z$ for the distances $s = 1 (a), 5 (b), 10 (c)$. Solid lines present calculations with the exact formula (27), dashed lines show approximations with Eq. (28).

Similarly, correlations of the $z$-component of the electric field are also cylindrically symmetrical and may be represented in the plane $\mathbf{s} = (s\cdot\sin(\vartheta), 0, s\cdot\cos(\vartheta))$,

$$r_{zz}(\mathbf{s},\tau) = B_{zz}(\sqrt{2\eta\tau})\int_0^\infty dk\, k^2 e^{-3k^2/2} \int_{-1}^1 du\, u^4 e^{-2\eta\tau u^2} \cos\left(\frac{kus\cos\vartheta}{L(\tau)}\right) J_0\left(\frac{k\sqrt{1-u^2}}{L(\tau)} s\sin\vartheta\right) \tag{29}$$

with the auxiliary function

$$B_{zz}(b) = \frac{4\sqrt{6}}{\pi}\frac{b^5}{\mathrm{erf}(b)}\left[1 - \frac{2}{\sqrt{\pi}}\frac{b\exp(-b^2)}{\mathrm{erf}(b)}\left(1 + \frac{2}{3}b^2\right)\right]^{-1}. \tag{30}$$

The development of $r_{zz}(\mathbf{s},\tau)$ as a function of $\vartheta$ at the sphere $s = 3$ exhibits a nonmonotonic behavior in time $\tau$ (Fig. 4). From the initial anisotropic state (*a*) with preferred correlations in the polarization direction it first develops to quite anisotropic transverse correlations (*b*,*c*) and then evolves back to rather isotropic strong correlations slightly preferred in the polarization direction. This can also be interpreted in terms of the relation between $s$ and the correlation length $L(\tau)$.

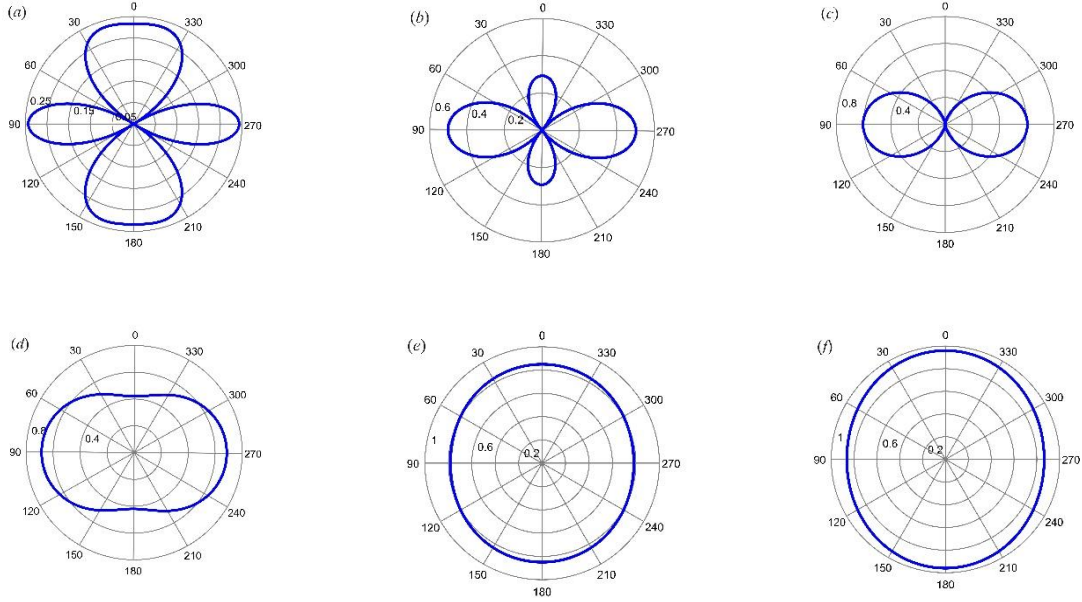

Fig. 4. Dependence of the correlation coefficient $r_{zz}(\mathbf{s},\tau)$ at the distance $s=3$ in the cross section $(s_x, 0, s_z)$ on the polar angle $\vartheta$ with respect to the polar axis $z$ for the moments $\tau = 0.01\ (a), 0.5\ (b), 1\ (c), 2\ (d),\ 5\ (e), 10\ (f)$.

The correlation coefficients for the $z$-components of the polarization, Fig. 2, and of the electric field, Fig. 4, exhibit remarkably different trends explained by the distinct physical properties of these fields. Firstly, the polarization in TGS has the only $z$-component and thus cannot rotate in space, while the electric field has all three components and can freely rotate. This makes possible a complicated interplay between the field correlation coefficients $r_{zz}(\mathbf{s},\tau)$, $r_{xx}(\mathbf{s},\tau)$, $r_{yy}(\mathbf{s},\tau)$, $r_{xz}(\mathbf{s},\tau)$, $r_{yz}(\mathbf{s},\tau)$, $r_{xy}(\mathbf{s},\tau)$. Secondly, a physical difference between the polarization and the electric field consists in the fact that the former can be either short-range or long-range correlated (as soon as a periodic domain structure is formed), while the latter is always long-range correlated. At the early stages (*a-d*) in Fig. 2, small domains of the size $L(\tau)$ are hardly correlated at the distance $s \gg L(\tau)$. Weak correlations along the polarization direction are of the local nature. In contrast, already at the very early stage, Fig. 4(*a*), field correlations are substantial due to long-range electrostatic fields. Particularly, the z-components of the electric field demonstrate long-range transverse correlations in Figs. 4(*a-c*), because the $\epsilon_z$ field component results from contributions of many domains including the far-away ones beyond the distance $L(\tau)$. When $L(\tau)$ substantially exceeds $s$ with increasing time, as in Fig. 2(*e,f*) and Fig. 4(*e,f*), the pairs of random points at the distance $s$ are mostly immersed in one domain where both the polarization and electric field behave coherently, so that the correlations become strong and increasingly isotropic.

The variation of $r_{zz}(\mathbf{s},\tau)$ with the distance $s$ at a fixed time $\tau = 0.1$ exhibits a nontrivial and nonmonotonic behavior (Fig. 5). Starting from bilateral-symmetrical anisotropic correlations at short distances (*a*) it gradually transforms to almost 4-fold angle dependence (*b*,*c*) and then gradually transforms to the asymptotic 6-fold dependence at large $s > L(\tau)$ (*d-f*), though at very low amplitudes.

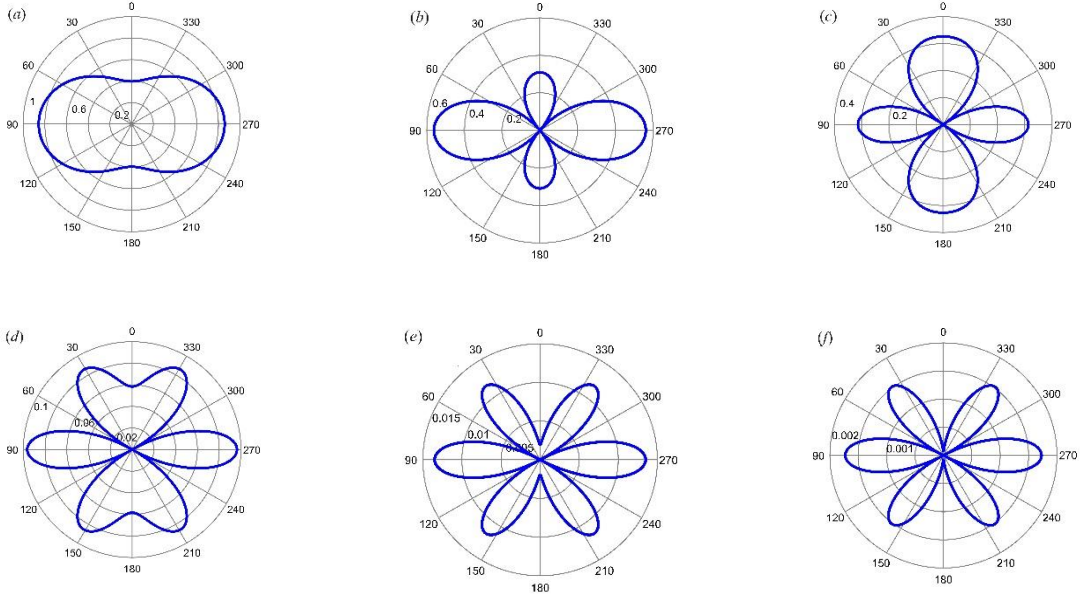

Fig. 5. Dependence of the correlation coefficient $r_{zz}(\mathbf{s},\tau)$ at the moment $\tau = 0.1$ on the polar angle $\vartheta$ with respect to the polar axis $z$ for the distances $s=1$ (*a*), 2 (*b*), 3 (*c*), 5 (*d*), 10 (*e*), 20 (*f*).

Due to the polarization-electric field coupling in the energy functional (1) the cross-correlation coefficient between *z*-components of polarization and field, $\psi_{zz}(\mathbf{s},\tau)$, plays a special role directly contributing to the evolution equations (10). It reads

$$\psi_{zz}(\mathbf{s},\tau) = C_{zz}\left(\sqrt{2\eta\tau}\right) \int_0^\infty dk\, k^2 e^{-3k^2/2} \int_{-1}^1 du\, u^2\, e^{-2\eta\tau u^2} \cos\left(\frac{kus_z}{L(\tau)}\right) J_0\left(\frac{ks_\perp\sqrt{1-u^2}}{L(\tau)}\right) \quad (31)$$

with the auxiliary function

$$C_{zz}(b) = -\frac{6\sqrt{2}}{\pi}\frac{b^3}{\mathrm{erf}(b)}\left[1 - \frac{2}{\sqrt{\pi}}\frac{b\exp(-b^2)}{\mathrm{erf}(b)}\left(1 + \frac{2}{3}b^2\right)\right]^{-1/2}. \quad (32)$$

From Eq. (31), the longitudinal correlation coefficient for the case with $\mathbf{s} = (0,0,s_z)$ can be calculated in a closed form, which can be used for testing numerical results,

$$\psi_{zz}^{\parallel}(s_z,\tau) = \frac{\pi}{6\sqrt{6}} C_{zz}(\sqrt{2\eta\tau}) \left[ \frac{\text{erf}\sqrt{2\eta\tau+v}}{(2\eta\tau+v)^{3/2}} \left(1 - \frac{3v}{2\eta\tau+v}\right) - \frac{2}{\sqrt{\pi}} \frac{\exp(-2\eta\tau-v)}{2\eta\tau+v} \left(1 - 2v - \frac{3v}{2\eta\tau+v}\right) \right], \quad (33)$$

where a combined variable $v = s_z^2/6L^2(\tau)$ was introduced for convenience.

The function $\psi_{zz}(\mathbf{s},\tau)$ also possesses a cylindrical symmetry allowing a presentation in the plane $\mathbf{s} = (s\cdot\sin(\vartheta), 0, s\cdot\cos(\vartheta))$. Its evolution with time at the sphere $s = 3$ (Fig. 6) reminds that of the function $r_{zz}(\mathbf{s},\tau)$ in Fig.4 for the same physical reasons. At the early stage (*a,b*) $\psi_{zz}(\mathbf{s},\tau)$ demonstrates very anisotropic correlations of bilateral symmetry pronounced in the polarization direction. This may indicate an initial tendency of the field-driven domain development in this direction. This trend is changed in the intermediate stage (*c,d*) to increasing transverse correlations which may indicate the field-driven transverse ordering of domains.

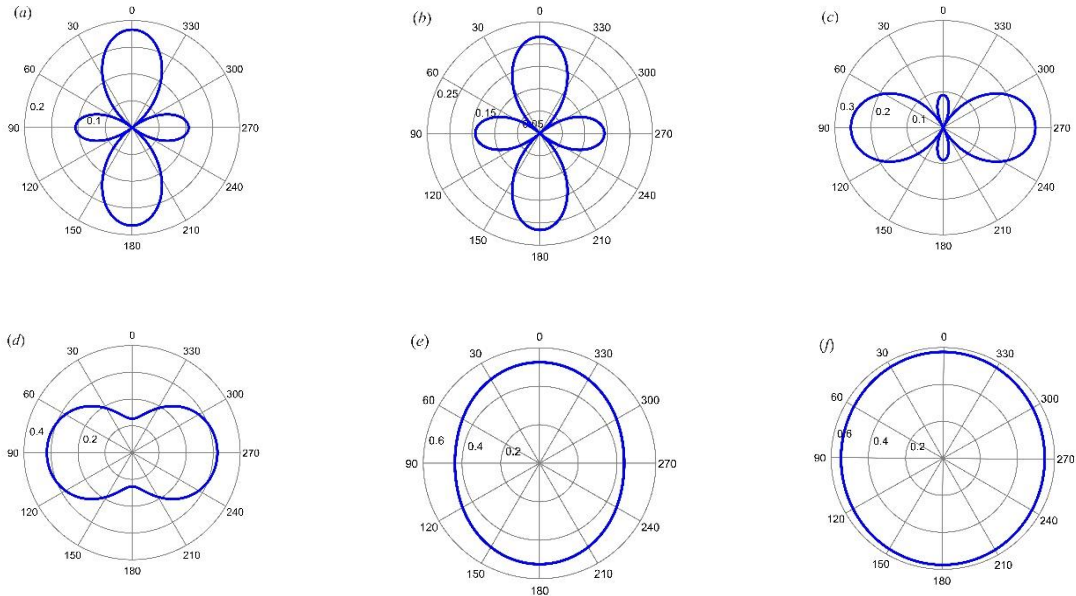

Fig. 6. Dependence of the correlation coefficient $\psi_{zz}(\mathbf{s},\tau)$ at the distance $s = 3$ on the polar angle $\vartheta$ with respect to the polar axis $z$ for the moments $\tau = 0.01\ (a), 0.1\ (b), 0.5\ (c), 1\ (d), 5\ (e), 30\ (f)$.

With the increasing correlation length $L(\tau)$ at the later stage (*e,f*) correlations become increasingly isotropic and strong, however, again slightly preferred in polarization direction.

The variation of $\psi_{zz}(\mathbf{s},\tau)$ with the distance $s$ at a fixed time $\tau$ again exhibits a nontrivial and nonmonotonic behavior (Fig. 7). Starting with substantial and virtually isotropic correlations at a short distance $s < L(\tau)$ (*a*) the function transforms to rather transverse correlations at intermediate distances $s \sim L(\tau)$ (*b,c*), then evolves to almost 4-fold correlations at $s > L(\tau)$ (*d*) to develop later to rather anisotropic though weak correlations preferably in polarization direction (*e,f*) at $s \gg L(\tau)$.

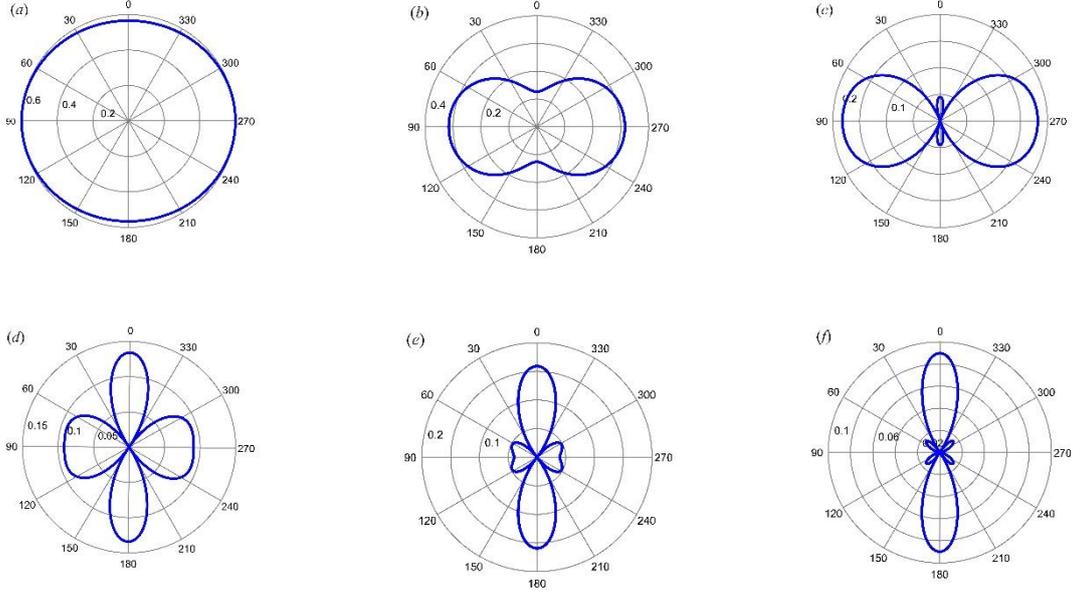

Fig. 7. Dependence of the correlation coefficient $\psi_{zz}(\mathbf{s},\tau)$ at the moment $\tau = 1$ on the polar angle $\vartheta$ with respect to the polar axis $z$ for the distances $s = 1$ (*a*), 3 (*b*), 4 (*c*), 5 (*d*), 6 (*e*), 10 (*f*).

### 4.2 Correlation coefficients breaking cylindrical symmetry

The correlation coefficient for the *x*-component of the electric field (17) is as follows,

$$r_{xx}(\mathbf{s},\tau) = B_{xx}(\sqrt{2\eta\tau}) \int_0^\infty dk\, k^2 e^{-3k^2/2} \int_{-1}^{1} du\, u^2(1-u^2) e^{-2\eta\tau u^2} \cos\left(\frac{kus_z}{L(\tau)}\right)$$

$$\times \int_0^{2\pi} d\phi \cos^2\phi \cos\left[\frac{k\sqrt{1-u^2}}{L(\tau)}(s_x\cos\phi + s_y\sin\phi)\right] \qquad (34)$$

with the auxiliary function

$$B_{xx}(b) = \frac{6\sqrt{6}}{\pi^2}\frac{b^3}{\operatorname{erf}(b)}\left[1 - \frac{3}{2b^2}\left(1 - \frac{2}{\sqrt{\pi}}\frac{b\exp(-b^2)}{\operatorname{erf}(b)}\right)\right]^{-1}. \qquad (35)$$

Its orientational dependence may be represented on a sphere of constant distance $s$ as is shown for example in Fig. 8 for $s = 3$ at the moment $\tau = 1$. It has a complicated 3D shape and develops in time from very anisotropic medium correlations towards stronger but more isotropic correlations somewhat in favor of polarization direction. The dependence on the distance $s$ exhibits nonmonotonic trends. Correlations are slightly anisotropic in favor of polarization direction at short distances, then develop rather in favor of transverse correlations, and finally, at large distances, remain weak and very anisotropic again in favor of polarization direction.

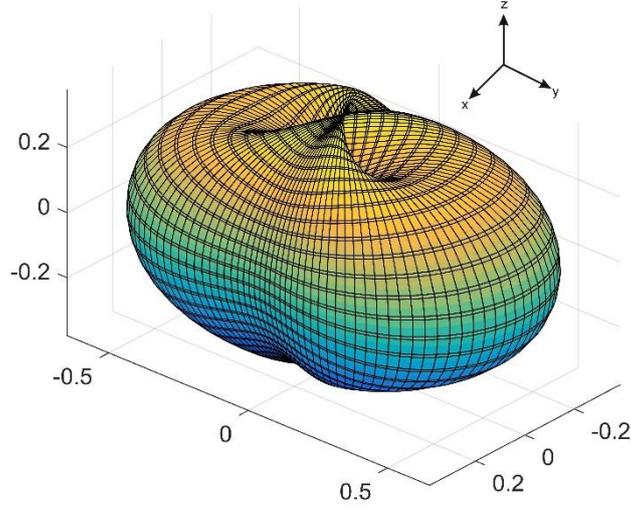

Fig. 8. Orientational dependence of the correlation coefficient $r_{xx}(\mathbf{s},\tau)$ at the moment $\tau = 1$ in spherical coordinates on the surface of the sphere $s = 3$.

Similarly to $r_{xx}$, for the *y*-component of the electric field (17),

$$r_{yy}(\mathbf{s},\tau) = B_{yy}(\sqrt{2\eta\tau}) \int_0^\infty dk\, k^2 e^{-3k^2/2} \int_{-1}^1 du\, u^2(1-u^2) e^{-2\eta\tau u^2} \cos\left(\frac{kus_z}{L(\tau)}\right)$$

$$\times \int_0^{2\pi} d\phi \sin^2\phi \cos\left[\frac{k\sqrt{1-u^2}}{L(\tau)}(s_x \cos\phi + s_y \sin\phi)\right], \qquad (36)$$

where the function $B_{yy}(b) = B_{xx}(b)$ for symmetry reasons. The orientational dependence of this function is similar to that of $r_{xx}$ but spatially rotated by the azimuthal angle $\pi/2$, therefore it is not presented graphically.

The cross-correlations of the electric field components are described by the coefficients (18). Particularly, $r_{xy}(\mathbf{s},\tau) = R_{xy}(\mathbf{s},\tau)/R_{xx}(\mathbf{0},\tau)$ because $R_{yy}(\mathbf{0},\tau) = R_{xx}(\mathbf{0},\tau)$. Thus, for a general direction of the vector $\mathbf{s} = (s_x, s_y, s_z)$,

$$r_{xy}(\mathbf{s},\tau) = B_{xy}(\sqrt{2\eta\tau}) \int_0^\infty dk\, k^2 e^{-3k^2/2} \int_{-1}^1 du\, u^2(1-u^2) e^{-2\eta\tau u^2} \cos\left(\frac{kus_z}{L(\tau)}\right)$$

$$\times \int_0^{2\pi} d\phi \cos\phi \sin\phi \cos\left[\frac{k\sqrt{1-u^2}}{L(\tau)}(s_x \cos\phi + s_y \sin\phi)\right] \qquad (37)$$

where $B_{xy}(b) = B_{xx}(b)$. We note that here the correlations vanish at the planes $s_x = 0$ and $s_y = 0$ due to integration over the azimuthal angle $\phi$ in Eq. (37) irrespectively of the form of the initial correlation function $\widetilde{K}(\mathbf{q},0)$ if it is isotropic in $q$-space. Particularly, $r_{xy}(\mathbf{s},\tau) = 0$ at the line $\mathbf{s} = (0,0,s_z)$.

The orientational dependence of $r_{xy}$ may be represented on a sphere of constant distance $s$ as is shown in Fig. 9 for $s = 3$ and $\tau = 1$. With increasing time $\tau \gg 1$ the shape of the figure remains but the amplitude of correlations gets reduced. At short times $\tau < 1$ the amplitude of correlations gets stronger and acquires multiple small anisotropic features.

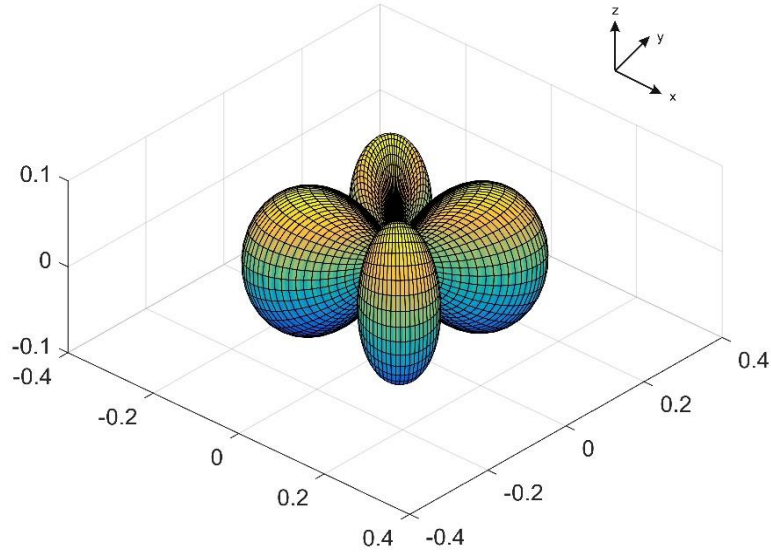

Fig. 9. Orientational dependence of the correlation coefficient $r_{xy}(\mathbf{s},\tau)$ at the moment $\tau = 1$ in spherical coordinates on the surface of the sphere $s = 3$.

The cross-correlation coefficient of the electric field components $r_{xz}(\mathbf{s},\tau)$ is given by

$$r_{xz}(\mathbf{s},\tau) = B_{xz}(\sqrt{2\eta\tau}) \int_0^\infty dk\, k^2 e^{-3k^2/2} \int_{-1}^1 du\, u^3 \sqrt{1-u^2}\, e^{-2\eta\tau u^2} \sin\left(\frac{kus_z}{L(\tau)}\right)$$

$$\times \int_0^{2\pi} d\phi \cos\phi \sin\left[\frac{k\sqrt{1-u^2}}{L(\tau)}(s_x \cos\phi + s_y \sin\phi)\right] \qquad (38)$$

with the auxiliary function

$$B_{xz}(b) = -\frac{12\,b^4}{\pi^2\sqrt{2}}\left[1 - \frac{3}{2b^2}\left(1 - \frac{2}{\sqrt{\pi}}\frac{b\exp(-b^2)}{\mathrm{erf}(b)}\right)\right]^{-1/2}\left[1 - \frac{2}{\sqrt{\pi}}\frac{b\exp(-b^2)}{\mathrm{erf}(b)}\left(1 + \frac{2}{3}b^2\right)\right]^{-1/2}. \qquad (39)$$

Similarly, the cross-correlation coefficient of the electric field components $r_{yz}(\mathbf{s},\tau)$ reads

$$r_{yz}(\mathbf{s},\tau) = B_{yz}(\sqrt{2\eta\tau}) \int_0^\infty dk\, k^2 e^{-3k^2/2} \int_{-1}^1 du\, u^3 \sqrt{1-u^2}\, e^{-2\eta\tau u^2} \sin\left(\frac{kus_z}{L(\tau)}\right)$$

$$\times \int_0^{2\pi} d\phi \sin\phi \sin\left[\frac{k\sqrt{1-u^2}}{L(\tau)}(s_x \cos\phi + s_y \sin\phi)\right] \qquad (40)$$

where $B_{yz}(b) = B_{xz}(b)$ for symmetry reasons. The coefficients $r_{xz}(\mathbf{s},\tau) = r_{yz}(\mathbf{s},\tau) = 0$ at the line $\mathbf{s} = (0,0,s_z)$ parallel to the polarization direction due to integration over $\phi$ in Eqs. (38,40) and they both also vanish at the plane $\mathbf{s} = (\mathbf{s}_\perp, 0)$ perpendicular to the polarization which is obvious from Eqs. (38,40). Furthermore, $r_{xz}$ vanishes at the plane $s_x = 0$ and $r_{yz}$ at the plane $s_y = 0$ when integrating over $\phi$ in Eqs. (38) and, respectively, (40). Since $r_{xz}$ is changing sign, the orientational dependence of the magnitude $|r_{xz}(\mathbf{s},\tau)|$ is presented in Fig. 10 for example on a sphere of constant distance $s = 3$. It remains topologically similar at all times and is decreasing asymptotically. The orientational dependence of $r_{yz}(\mathbf{s},\tau)$ is similar but spatially rotated by the azimuthal angle $\pi/2$, therefore it is not presented graphically.

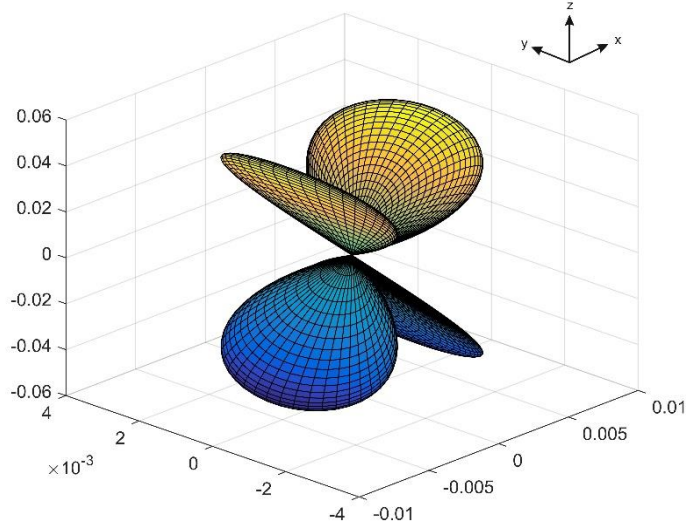

Fig. 10. Orientational dependence of the correlation coefficient $|r_{xz}(\mathbf{s},\tau)|$ at the moment $\tau = 1$ in spherical coordinates on the surface of the sphere $s = 3$.

Now we consider the cross-correlations between polarization and transverse field components. Particularly, the coefficient $\psi_{xz}$ reads

$$\psi_{xz}(\mathbf{s},\tau) = C_{xz}(\sqrt{2\eta\tau}) \int_0^\infty dk\, k^2 e^{-3k^2/2} \int_{-1}^1 du\, u\sqrt{1-u^2}\, e^{-2\eta\tau u^2} \sin\left(\frac{kus_z}{L(\tau)}\right)$$

$$\times \int_0^{2\pi} d\phi \cos\phi \sin\left[\frac{k\sqrt{1-u^2}}{L(\tau)}(s_x \cos\phi + s_y \sin\phi)\right] \qquad (41)$$

with the auxiliary function

$$C_{xz}(b) = -\frac{3\sqrt{6}\,b}{\pi^2\,\text{erf}(b)}\left[1 - \frac{3}{2b^2}\left(1 - \frac{2}{\sqrt{\pi}}\frac{b\exp(-b^2)}{\text{erf}(b)}\right)\right]^{-1/2}. \qquad (42)$$

Similarly,

$$\psi_{yz}(\mathbf{s},\tau) = C_{yz}(\sqrt{2\eta\tau})\int_0^\infty dk\,k^2 e^{-3k^2/2}\int_{-1}^1 du\,u\sqrt{1-u^2}\,e^{-2\eta\tau u^2}\sin\left(\frac{kus_z}{L(\tau)}\right)$$

$$\times \int_0^{2\pi} d\phi\,\sin\phi\,\sin\left[\frac{k\sqrt{1-u^2}}{L(\tau)}(s_x\cos\phi + s_y\sin\phi)\right] \qquad (43)$$

where $C_{yz}(b) = C_{xz}(b)$. Both correlation coefficients $\psi_{xz}(\mathbf{s},\tau)$ and $\psi_{yz}(\mathbf{s},\tau)$ vanish at the line $\mathbf{s} = (0,0,s_z)$ and in the plane $\mathbf{s} = (s_\perp, 0)$ as well as the coefficients $r_{xz}(\mathbf{s},\tau)$ and $r_{yz}(\mathbf{s},\tau)$ before. Coefficient $\psi_{xz}$ also vanishes at the plane $s_x = 0$ and $\psi_{yz}$ at the plane $s_y = 0$ when integrating over $\phi$ in Eqs. (41) and, respectively, (43). Since $\psi_{xz}$ is changing sign, the orientational dependence of the magnitude $|\psi_{xz}(\mathbf{s},\tau)|$ is presented in Fig. 11 for example at $s = 3$. The orientational dependence of $\psi_{yz}(\mathbf{s},\tau)$ is similar but spatially rotated by the azimuthal angle $\pi/2$, therefore it will not be presented graphically.

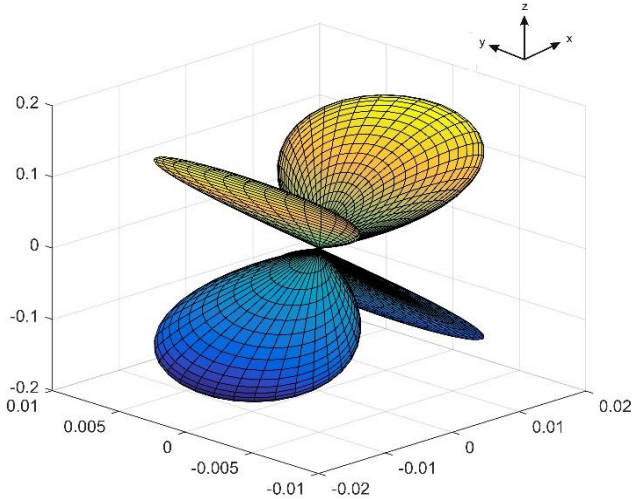

Fig. 11. Orientational dependence of the correlation coefficient $|\psi_{xz}(\mathbf{s},\tau)|$ at the moment $\tau = 1$ in spherical coordinates on the surface of the sphere $s = 3$.

Apparently, the shapes of $r_{xz}(\mathbf{s},\tau)$ and $\psi_{xz}(\mathbf{s},\tau)$ in Figs. 10 and 11 are pretty similar to each other that is not surprising, since the formulas (38) and (41) differ only in the power of the variable $u$ in addition to different time-dependent factors. Thus, they distinguish only quantitatively but not qualitatively.

## 5. Discussion

Account of fluctuations and correlations of random variables is necessary for description of thermodynamics of macroscopic random systems and their response to external fields. Thus, the knowledge of the dispersions (variances) of the polarization and electric field components as well as of the polarization correlations and the covariance $\langle P_z E_z \rangle$ is required to evaluate the energy of the stochastic domain system represented by Eq. (1).

Contribution of the polarization-field correlations represented in evolution equations (10) by the cross-correlation function $\widetilde{\Psi}_{zz}(\mathbf{q}, \tau) \sim \frac{q_z^2}{q^2} \widetilde{K}(\mathbf{q}, \tau)$ appears to be deciding for the formation of the multi-domain polarization state and has a great impact on the field-driven evolution of the domain system and the magnitude of the coercive field [52]. The time-dependent development of the correlation coefficient $\psi_{zz}(\mathbf{s}, \tau)$ presented in Fig. 6 gives an insight into the evolution of the domain structure pronouncing the formation of longitudinal individual domains at the early stage and of the transversely ordered domain structure at the later stage.

Spatial correlations become significant wherever nonlinear effects are involved and observations cover a large number of domains. Thus, the analysis of nonlinear piezoelectric response observed by PFM in ferroelectric films revealed medium- to large-scale correlations whose physical origin was not clearly understood [46,47]. Another example of nonlinear response is given by omnipresent electrostriction which is typically averaged over macroscopic areas and volumes requiring an account of fluctuations of random variables in disordered systems. The electrostriction effect quadratic in the electric field components may be affected by the electric field correlations represented by the correlation coefficients $r_{xx}$, Eq. (34), $r_{yy}$, Eq. (36), $r_{zz}$, Eq. (29), $r_{xy}$, Eq. (37), $r_{xz}$, Eq. (38), $r_{yz}$, Eq. (40). One more case for the implementation of the field correlations is given by the quadratic electro-optic (Kerr) effect in noncentrosymmetric crystals, such as ferroelectrics [62,63].

As was shown by Dolino [64], the knowledge of the correlation function of polarization $K(\mathbf{s}, \tau)$ is required for evaluation of the intensity of the second-harmonic scattering (SHS) in nonlinear ferroelectric crystals. This kind of response is quadratic in polarization and nonlocal, since it involves polarization values at different locations. The angular pattern of scattered light is dependent upon domain shapes and structures so that the variation of the second-harmonic intensity allows to judge on the changes in the domain structure. SHS as well as second harmonic generation (SHG) were successfully used to study the properties of crystals undergoing structural phase transitions [60,65].

Coming back to the fundamental question on the role of polarization-field correlations during the domain structure formation or the field-driven switching, we would like to remind on the earlier simulations, predicting very small or vanishing polarization-field cross-correlations in polycrystalline ferroelectrics [39,40]. Comparing these results with the results of this study obtained on the basis of the exactly solvable stochastic model for a uniaxial single-crystalline ferroelectric [52] we note that some cross-correlations in a single crystal indeed vanish at certain directions and planes. For example, the coefficients $r_{xz}(\mathbf{s}, \tau)$ and $\psi_{xz}(\mathbf{s}, \tau)$ vanish at the planes $s_x = 0$ and $s_z = 0$, while the coefficient $r_{xy}(\mathbf{s}, \tau)$ vanishes at the planes $s_x = 0$ and $s_y = 0$. Actually, coefficients $r_{xz}(\mathbf{s}, \tau)$, $r_{yz}(\mathbf{s}, \tau)$, $\psi_{xz}(\mathbf{s}, \tau)$ and $\psi_{yz}(\mathbf{s}, \tau)$ are negligible everywhere but a few narrow orientational regions seen in Figs. 10 an 11 where they are also small, and this is not accidental.

In fact, there are fundamental physical reasons for the disappearance of cross-correlations. They are expressed by the Obukhov theorem stating the absence of correlations between isotropic potential and solenoidal fields [66]. Electric field in our problem is potential one. Polarization field would be solenoidal if $\nabla \cdot \boldsymbol{P}$ were 0, i. e. if polarization would only build domain structures compensating all bound charges. This tendency was apparently realized in multiaxial polycrystalline ferroelectrics which were simulated in our SMS model [39,40], leading thus to the disappearance of cross-correlations. The tendency to form only charge-free domain walls under dynamic switching conditions was observed in phase-field [36] and molecular dynamics simulations [38] as well as in experiments [42,43]. However, the condition $\nabla \cdot \boldsymbol{P} = \boldsymbol{0}$ cannot be fully realized in a uniaxial ferroelectric like TGS. Nevertheless, the tendency to build charge-free polarization configurations like a quasi-periodic stripe structures prevails also in TGS. Thus, the cross-correlations exhibit, in principle, the tendency to vanish, which cannot be fully realized, so that such correlations remain in some directions. This statement applies to the cross-correlations $\psi_{xz}$, Eq. (41) and $\psi_{yz}$, (43), exhibiting non-vanishing amplitudes in some narrow spatial angle regions, Fig. 11. The cross-correlation coefficient of the different field components, like $r_{xz}$ Eq. (38) appears to have a structure quite similar to that of $\psi_{xz}$ and thus possess very similar properties represented by Fig. 10. Therefore, the disappearance (at least partial) of a range of cross-correlations has fundamental physical causes related to the potential and, respectively, solenoidal nature of the involved physical fields. However, differently from the multiaxial polycrystalline ferroelectrics [39,40], the $\psi_{zz}$ correlations between *z*-components of polarization and field undoubtedly play an important role in the formation of the domain structure in uniaxial single crystals and are not small.

We note that the correlation coefficients (16-18) describe correlation functions normalized to the variance of polarization, $D(\tau)$. If the system evolves towards a single-domain state, that occurs under applied fields higher than the coercive one [52], the polarization dispersion $D(\tau) \to 0$ asymptotically and all absolute values of fluctuations and correlations vanish simultaneously. Nevertheless, in this case correlations play an important role over the whole stage of the domain structure building too.

## 6. Conclusions

In order to fully understand the role of correlations in the process of domain formation and electric field-driven switching, both the temporal and spatial correlations between the emerging polarization and depolarization fields have to be accounted. This challenging task could not yet be carried out in full within the stochastic approach. The temporal correlations were partly considered by the multistep switching mechanism model [33-35] where the spatial correlations were still not taken into account. In the current study, we investigated the spatial correlations between emerging polarization domains and depolarization fields developing in time but still could not account for the mutual influence of polarization domains appearing at different times and different locations, which remains a task for the future. Nevertheless, the analytical results for the correlation coefficients between all involved stochastic variables, obtained on the basis of the exactly solvable stochastic model of polarization development [52], allowed insight in the temporal development of anisotropic spatial characteristics of the emerging domain structure.

Unfortunately, the possibilities to compare our analytical results with experiment or simulations are currently limited. The predicted temporal development of the correlation length and the polarization correlation coefficient in TGS were successfully compared with experiments in Ref. [52]. However, for the comparison of our results on the 3D correlation coefficients the statistically analyzed 3D data are required. Though 3D simulations of ferroelectrics by means of the phase-field approach, molecular dynamics and Monte Carlo method have been known for some time, we are not aware of the statistical analysis of the temporal simulation data. Particularly, the temporal 3D simulations of the domain formation in TGS are not available. To the best of our knowledge, there is just one experimental work by Wehmeier et al. presenting time-dependent 3D data on the domain structure formation in TGS obtained by second-harmonic generation microscopy [60], but unfortunately without statistical data processing. Therefore, our calculations of the correlation coefficients are primarily predictive. Hopefully, they will encourage researchers to carry out statistical processing of the available 3D data from experiments and simulations.

## Acknowledgements

O.Y.M. is grateful for the financial support by the Marie Sklodowska-Curie Actions for Ukraine (№ 1233427). This work was supported by the Deutsche Forschungsgemeinschaft (German Research Society, DFG) via Grant No. 405631895 (GE-1171/8-1).

## Appendix A: Relations between the correlation functions

Let us consider the Gauss equation equivalent to the Poisson equation (3b) taken at the point $\mathbf{r}_2$

$$\frac{\partial \epsilon_x(\mathbf{r}_2,\tau)}{\partial X_2} + \frac{\partial \epsilon_y(\mathbf{r}_2,\tau)}{\partial Y_2} + \frac{\partial \epsilon_z(\mathbf{r}_2,\tau)}{\partial Z_2} = -\eta \frac{\partial \xi(\mathbf{r}_2,\tau)}{\partial Z_2} \tag{A1}$$

and multiply it with $\epsilon_x(\mathbf{r}_1,\tau)$. After subsequent statistical averaging this leads to equation

$$\frac{\partial}{\partial X_2} \langle \epsilon_x(\mathbf{r}_1,\tau)\epsilon_x(\mathbf{r}_2,\tau) \rangle + \frac{\partial}{\partial Y_2} \langle \epsilon_x(\mathbf{r}_1,\tau)\epsilon_y(\mathbf{r}_2,\tau) \rangle + \frac{\partial}{\partial Z_2} \langle \epsilon_x(\mathbf{r}_1,\tau)\epsilon_z(\mathbf{r}_2,\tau) \rangle =$$

$$-\eta \frac{\partial}{\partial Z_2} \langle \epsilon_x(\mathbf{r}_1,\tau)\xi(\mathbf{r}_2,\tau) \rangle \tag{A2}$$

By substituting the field expression (2) into the latter equation its first term is transformed to

$$\frac{\partial}{\partial X_2} \langle \frac{\partial}{\partial X_1} \phi(\mathbf{r}_1,\tau) \frac{\partial}{\partial X_2} \phi(\mathbf{r}_2,\tau) \rangle = \frac{\partial}{\partial X_1} \frac{\partial^2}{\partial X_2^2} g(\mathbf{s},\tau). \tag{A3}$$

with $\mathbf{s} = \mathbf{r}_1 - \mathbf{r}_2$. Similarly, the second term is transformed to

$$\frac{\partial}{\partial Y_2} \langle \frac{\partial}{\partial X_1} \phi(\mathbf{r}_1,\tau) \frac{\partial}{\partial Y_2} \phi(\mathbf{r}_2,\tau) \rangle = \frac{\partial}{\partial X_1} \frac{\partial^2}{\partial Y_2^2} g(\mathbf{s},\tau) \tag{A4}$$

and the third term to

$$\frac{\partial}{\partial Z_2} \langle \frac{\partial}{\partial X_1} \phi(\mathbf{r}_1,\tau) \frac{\partial}{\partial Z_2} \phi(\mathbf{r}_2,\tau) \rangle = \frac{\partial}{\partial X_1} \frac{\partial^2}{\partial Z_2^2} g(\mathbf{s},\tau). \tag{A5}$$

The right-hand side of Eq. (A2) represents a derivative of $\Psi_{xz}(\mathbf{s},\tau)$. By substituting the relations (A3-A5) into Eq. (A2) one obtains

$$\frac{\partial}{\partial s_x} \Delta g(\mathbf{s},\tau) = \eta \frac{\partial}{\partial s_z} \Psi_{xz}(\mathbf{s},\tau) \tag{A6}$$

with the Laplace operator $\Delta$.

Similarly, by multiplying Eq. (A1) with $\epsilon_y(\mathbf{r_1}, \tau)$ with subsequent statistical averaging one obtains equation

$$\frac{\partial}{\partial s_y}\Delta g(\mathbf{s}, \tau) = \eta \frac{\partial}{\partial s_z} \Psi_{yz}(\mathbf{s}, \tau) \qquad (A7)$$

and by multiplying Eq. (A1) with $\epsilon_z(\mathbf{r_1}, \tau)$ with subsequent statistical averaging one obtains equation

$$\frac{\partial}{\partial s_z}\Delta g(\mathbf{s}, \tau) = \eta \frac{\partial}{\partial s_z} \Psi_{zz}(\mathbf{s}, \tau). \qquad (A8)$$

Considering vanishing of correlations at infinity, the latter relation means that

$$\Delta g(\mathbf{s}, \tau) = \eta \Psi_{zz}(\mathbf{s}, \tau). \qquad (A9)$$

Now, by substituting Eq. (A9) into Eq. (A6) one obtains

$$\frac{\partial}{\partial s_x} \Psi_{zz}(\mathbf{s}, \tau) = \frac{\partial}{\partial s_z} \Psi_{xz}(\mathbf{s}, \tau) \qquad (A10)$$

and by substituting Eq. (A9) into Eq. (A7) one obtains

$$\frac{\partial}{\partial s_y} \Psi_{zz}(\mathbf{s}, \tau) = \frac{\partial}{\partial s_z} \Psi_{yz}(\mathbf{s}, \tau). \qquad (A11)$$

Eqs. (A9-A11) provide together the relations (4).